\documentclass[aps,pra,groupedaddress]{revtex4}
\usepackage{graphics}
\usepackage{amsmath}
\usepackage{graphicx}
\usepackage{amsfonts}
\usepackage{amssymb}

\begin{document}

\title{Symmetric collective attacks for the eavesdropping of symmetric
quantum key distribution}
\author{Stefano Pirandola}
\affiliation{M.I.T. - Research Laboratory of Electronics,
Cambridge MA 02139, USA}

\begin{abstract}
We consider the collective eavesdropping of the BB84 and six-state
protocols. Since these protocols are symmetric in the eigenstates
of conjugate bases, we consider collective attacks having the same
kind of symmetry. We then show how these symmetric collective
attacks are sufficiently strong in order to minimize the
Devetak-Winter rates. In fact, it is quite easy to construct
simple examples able to reach the unconditionally-secure key-rates
of these protocols.
\end{abstract}

\maketitle

\section{Introduction}

Recently, Renner \cite{Renato} has shown how to reduce quantum key
distribution (QKD) to the cryptoanalysis of collective attacks.
This is possible by turning an arbitrary QKD protocol into a
permutation invariant one, where Alice and Bob publicly agree on a
random permutation which they use to reorder their classical
values just at the end of the quantum communication and before any
other classical processing of the data \cite{Renato}. Thanks to
this permutation invariance, a finite quantum de Finetti theorem
\cite{deFinetti} can be applied to the cryptographic scenario and,
therefore, the most general coherent attack can be approximated by
a\emph{\ mixture} of collective attacks. As a consequence, a bound
on the key-rate for all the possible collective attacks becomes
automatically a bound for the most general attacks allowed by
quantum
mechanics. Since a natural upper bound for the eavesdropper's information $%
I_{AE}$ is given by the Holevo information $\chi _{AE}$, the minimization of
the Devetak-Winter rate \cite{DeveWinterRATE,NoteDWrate} $%
R_{DW}:=I_{AB}-\chi _{AE}$ on the class of collective attacks
provide a natural lower bound for the unconditionally-secure
key-rate.

In this paper we consider the cryptoanalysis of the BB84 and the
six-state protocols. Such QKD schemes can be called
\emph{symmetric} since they are based on the symmetric
exploitation of the eigenstates of conjugate bases (mutually
unbiased bases). It is then intuitive to consider collective
attacks whose action is symmetric on these eigenstates, resulting
in uniform contractions within the Bloch sphere. Such symmetric
collective attacks are in fact a trivial extension of the
symmetric individual attacks defined by Gisin \textit{et al.}
\cite{Symm}. The naive result of this paper is that, for symmetric
QKD schemes like the BB84 and six-state protocols, the
minimization of the Devetak-Winter rates can be restricted to the
class of symmetric collective attacks. In fact, it is very easy to
find simple examples of symmetric collective attacks whose
Devetak-Winter rates correspond exactly to the
unconditionally-secure key-rates of these symmetric protocols.

\section{The BB84 protocol and its symmetric eavesdropping}

In the BB84 protocol \cite{BB84}, two honest users (Alice and Bob) randomly
choose between two conjugate bases, i.e., the $Z$-basis $\{\left\vert
0\right\rangle ,\left\vert 1\right\rangle \}$ (the eigenstates of the Pauli
operator $Z$) and the $X$-basis $\{\left\vert +\right\rangle ,\left\vert
-\right\rangle \}$ (the eigenstates $\left\vert \pm \right\rangle
=2^{-1/2}\left( \left\vert 0\right\rangle \pm \left\vert 1\right\rangle
\right) $ of the Pauli operator $X$). Alice encodes a logical bit into her
basis $\sigma _{A}=Z\vee X$ according to the mapping $\bar{0}=\left\vert
0\right\rangle \vee \left\vert +\right\rangle $ and $\bar{1}=\left\vert
1\right\rangle \vee \left\vert -\right\rangle $. The signal state $%
\left\vert u\right\rangle $ with $u=\{0,1,+,-\}$ is then sent to Bob through
the noisy channel $\mathcal{E}$, who will project the output state $\rho
_{B}(u):=\mathcal{E}\left( \left\vert u\right\rangle \left\langle
u\right\vert \right) $ onto his basis $\sigma _{B}=Z\vee X$ in order to
decode Alice's logical bit. At the end of the quantum communication, Alice
and Bob publicly agree a random permutation of their binary data (called the
\emph{raw key}). Then, they disclose all their bases (\emph{basis
reconciliation}) and keep only the compatible data, forming the so-called
\emph{sifted key}. Such a key is still affected by errors due to the noise
of the channel and the corresponding error rate is called QBER (for \emph{%
quantum bit error rate}). The QBER is computed during the subsequent \emph{%
error estimation}, where the honest users publicly compare a (small) random
subset of the sifted key. From the knowledge of the QBER, the honest users
can bound the amount of information potentially stolen by an eavesdropper
(Eve). In particular, if the QBER is below a certain security threshold,
then Alice and Bob can apply procedures of error correction and privacy
amplification in order to derive a final secret and error-free binary key.

In a collective attack, Eve probes each signal qubit using a fresh ancilla,
which is then stored in a cell of a quantum memory coherently measured at
the end of the protocol. In particular, such a coherent measurement is also
optimized on every classical communication used by Alice and Bob during the
protocol like, e.g., the basis reconciliation. As a consequence, Eve has an
\emph{a posteriori }knowledge of the basis ($Z$ or $X$) which was used for
each signal qubit. On the one hand, Eve can exploit this knowledge in the
final detection \cite{CoheDET}, on the other hand, she cannot exploit it for
a conditional optimization of the signal-ancilla interactions (which, of
course, have been already occurred). Since the usage of the two conjugate
bases is perfectly symmetric in the BB84 protocol, the optimal eavesdropping
strategy should consist in signal-ancilla interactions which are symmetric
in the eigenstates of these conjugate bases.

Let us explicitly construct this kind of symmetric interaction. According to
the Stinespring dilation theorem \cite{Stine}, the quantum channel $\mathcal{%
E}$ acting on the signal qubit can be represented by a unitary interaction $%
\hat{U}$ coupling the signal qubit with two ancillary qubits initially
prepared in the vacuum state (such a representation is also minimal and
unique up to partial isometries). Then, for every input $u=\{0,1,+,-\}$, we
can write the following signal-ancilla unitary interaction%
\begin{equation}
\hat{U}\left( \left\vert u\right\rangle \otimes \left\vert 0,0\right\rangle
\right) =\left\vert u\right\rangle \left\vert F_{u}\right\rangle +\left\vert
u\oplus 1\right\rangle \left\vert D_{u}\right\rangle ~,  \label{U_gen}
\end{equation}%
where $u\oplus 1=\{1,0,-,+\}$ and the output ancillas ($F$ and $D$'s) are
generally not orthogonal neither normalized. Now, the condition of symmetry
in the four eigenstates $\left\vert u\right\rangle $ reduces the number of
possible unitaries $\hat{U}$. In particular, by imposing the conditions%
\begin{equation}
\left\langle F_{u}\right. \left\vert F_{u}\right\rangle =F~,~\left\langle
D_{u}\right. \left\vert D_{u}\right\rangle =D:=1-F~,~\left\langle
F_{u}\right. \left\vert D_{u}\right\rangle =0~,  \label{symmetric}
\end{equation}%
one makes $\hat{U}$ symmetric and Eq.~(\ref{U_gen}) a Schmidt form. The
Stinespring dilation of Eq.~(\ref{U_gen}) under the conditions of Eq.~(\ref%
{symmetric}) defines the notion of symmetric attack, which is then
individual or collective depending on the kind of measurement performed by
Eve on her quantum memory. The corresponding action on the Alice-Bob channel
is given by the map%
\begin{equation}
\mathcal{E}:\left\vert u\right\rangle \left\langle u\right\vert \rightarrow
\rho _{B}(u)=F\left\vert u\right\rangle \left\langle u\right\vert
+D\left\vert u\oplus 1\right\rangle \left\langle u\oplus 1\right\vert ~,
\label{Symmetric_Channel}
\end{equation}%
describing a uniform contraction by $F-D$ of the signal states, which is
here equivalent to the contraction of the equator of the Bloch sphere \cite%
{Symm}. From Eq.~(\ref{Symmetric_Channel}) it is clear that parameter $F$
represents the fidelity while $D$ is the QBER. As a consequence, Alice and
Bob's mutual information is simply given by $I_{AB}=1-H(D)$ where $%
H(p)=-p\log _{2}p-(1-p)\log _{2}(1-p)$ is the Shannon entropy.

Let us now consider the output state $\rho _{E}(u)$ which is received by Eve
in the complementary Alice-Eve channel $\mathcal{\tilde{E}}:\left\vert
u\right\rangle \left\langle u\right\vert \rightarrow \rho _{E}(u)$. This is
equal to%
\begin{equation}
\rho _{E}(u)=\left\vert F_{u}\right\rangle \left\langle F_{u}\right\vert
+\left\vert D_{u}\right\rangle \left\langle D_{u}\right\vert =F\left\vert
f_{u}\right\rangle \left\langle f_{u}\right\vert +D\left\vert
d_{u}\right\rangle \left\langle d_{u}\right\vert ~,  \label{outputEVE}
\end{equation}%
where the normalized states $\left\vert f_{u}\right\rangle
:=F^{-1/2}\left\vert F_{u}\right\rangle $ and $\left\vert d_{u}\right\rangle
:=D^{-1/2}\left\vert D_{u}\right\rangle $ have been introduced. In case of
collective attack, this output state is subject to an optimal coherent
measurement involving all the cells of the quantum memory. Since Eve has the
a posteriori knowledge of the basis, her coherent measurement has to
discriminate between the two states of the quantum ensemble%
\begin{equation}
\mathcal{Q}=\left\{
\begin{array}{l}
\rho _{E}(u)~,~p(u)=\frac{1}{2} \\
\\
\rho _{E}(u\oplus 1)~,~p(u\oplus 1)=\frac{1}{2}%
\end{array}%
\right. ~\Rightarrow ~\rho _{E}:=\frac{\rho _{E}(u)+\rho _{E}(u\oplus 1)}{2}%
~.  \label{Eve_Ensemble}
\end{equation}%
It is known that the maximal amount of classical information (\emph{%
accessible information}) that Eve can steal from this ensemble is
upper-bounded by the Holevo information%
\begin{equation}
\chi _{AE}:=S(\rho _{E})-\frac{S[\rho _{E}(u)]+S[\rho _{E}(u\oplus 1)]}{2}~,
\end{equation}%
where $S(\rho ):=-\mathrm{Tr}\left( \rho \log _{2}\rho \right) $ is the Von
Neumann entropy. As a consequence, the secret-key rate is lower bounded by
the Devetak-Winter rate \cite{NoteDWrate}%
\begin{equation}
R_{DW}:=I_{AB}-\chi _{AE}~.
\end{equation}%
Since $\left\langle f_{u}\right. \left\vert d_{u}\right\rangle =\left\langle
F_{u}\right. \left\vert D_{u}\right\rangle =0$ in Eq.~(\ref{outputEVE}), we
have that $S[\rho _{E}(u)]=S[\rho _{E}(u\oplus 1)]=H(D)$. By exploiting this
expression and $I_{AB}=1-H(D)$, the Devetak-Winter rate for a symmetric
collective attack simply becomes
\begin{equation}
R_{DW}=1-S(\rho _{E})~,  \label{DWrateSimple}
\end{equation}%
where only $S(\rho _{E})$ remains to be computed.

Following Gisin \textit{et al}. \cite{Symm}, let us simplify the structure
of the symmetric attack by imposing the additional conditions%
\begin{equation}
\left\langle F_{u}\right. \left\vert F_{u\oplus 1}\right\rangle =F\cos
x~,~\left\langle D_{u}\right. \left\vert D_{u\oplus 1}\right\rangle =D\cos
y~,~\left\langle F_{u}\right. \left\vert D_{u\oplus 1}\right\rangle =0~,
\label{mixedTerms}
\end{equation}%
with $x$ and $y$ real numbers. Such conditions imply%
\begin{equation}
D=\frac{1-\cos x}{2-\cos x+\cos y}:=D(x,y)~,  \label{Fid}
\end{equation}%
so that $\hat{U}$ is not only symmetric but also completely determined by
two angles $x$ and $y$. In particular, we can realize all the conditions in
Eqs.~(\ref{symmetric}) and~(\ref{mixedTerms}) by choosing in Eq.~(\ref{U_gen}%
) the ancilla states \cite{Symm}%
\begin{equation}
\left\vert F_{0}\right\rangle =\left(
\begin{array}{c}
\sqrt{F} \\
0 \\
0 \\
0%
\end{array}%
\right) ~,~\left\vert D_{0}\right\rangle =\left(
\begin{array}{c}
0 \\
\sqrt{D} \\
0 \\
0%
\end{array}%
\right) ~,~\left\vert F_{1}\right\rangle =\left(
\begin{array}{c}
\sqrt{F}\cos x \\
0 \\
0 \\
\sqrt{F}\sin x%
\end{array}%
\right) ~,~\left\vert D_{1}\right\rangle =\left(
\begin{array}{c}
0 \\
\sqrt{D}\cos y \\
\sqrt{D}\sin y \\
0%
\end{array}%
\right) ~.  \label{choice}
\end{equation}%
Let us denote by $\mathcal{S}(x,y)$ the symmetric collective attack
specified by the interaction of Eq.~(\ref{choice}). Then, it is easy to
prove that the attack $\mathcal{S}(x,x)$ has a Devetak-Winter rate equal to
\begin{equation}
R_{DW}=1-2H(D)~,  \label{DWrateBB84}
\end{equation}%
which corresponds exactly to the unconditionally-secure key-rate
of the BB84 protocol \cite{BB84security} (with unconditional
security threshold $D\simeq 11\%$ as given by $1-2H(D)=0$).

\bigskip

\textbf{Proof of Eq.~(\ref{DWrateBB84}).~}\ In order to prove the rate of
Eq.~(\ref{DWrateBB84}) we have to compute the entropy $S(\rho _{E})$ in Eq.~(%
\ref{DWrateSimple}) by exploiting the properties of the attack $\mathcal{S}%
(x,x)$, which are simply given by conditions of Eqs.~(\ref{symmetric}) and~(%
\ref{mixedTerms}) with $x=y$. By introducing the states%
\begin{equation}
\rho _{F}:=\frac{1}{2}\left( \left\vert f_{u}\right\rangle \left\langle
f_{u}\right\vert +\left\vert f_{u\oplus 1}\right\rangle \left\langle
f_{u\oplus 1}\right\vert \right) ~,~\rho _{D}:=\frac{1}{2}\left( \left\vert
d_{u}\right\rangle \left\langle d_{u}\right\vert +\left\vert d_{u\oplus
1}\right\rangle \left\langle d_{u\oplus 1}\right\vert \right) ~,
\label{rho_di}
\end{equation}%
we can recast the average state $\rho _{E}$ of Eq.~(\ref{Eve_Ensemble}) in
the form $\rho _{E}=F\rho _{F}+D\rho _{D}$, so that it can be equivalently
seen as the average state of the quantum ensemble%
\begin{equation}
\mathcal{\tilde{Q}}=\left\{
\begin{array}{c}
\rho _{F}~,~p(F)=F \\
\rho _{D}~,~p(D)=D%
\end{array}%
\right. ~.  \label{Q_tilde}
\end{equation}%
From Eqs.~(\ref{symmetric}) and~(\ref{mixedTerms}) we easily derive that $%
\rho _{F}$ and $\rho _{D}$ are orthogonal, i.e., $\mathrm{Tr}(\rho _{F}\rho
_{D})=0$. As a consequence, we have%
\begin{equation}
\chi (\mathcal{\tilde{Q}}):=S(\rho _{E})-[FS(\rho _{F})+DS(\rho _{D})]=H(D)~.
\label{Extract}
\end{equation}%
In order to extract $S(\rho _{E})$ from Eq.~(\ref{Extract}), we have to
compute the two entropies $S(\rho _{F})$ and $S(\rho _{D})$. For computing $%
S(\rho _{F})$ let us introduce the orthonormal set $\{\left\vert
f_{u}\right\rangle ,\left\vert f_{u}^{\perp }\right\rangle \}$, where $%
\left\vert f_{u}^{\perp }\right\rangle $ is an arbitrary vector defined by $%
\left\langle f_{u}\right. \left\vert f_{u}^{\perp }\right\rangle =0$ and $%
\left\langle f_{u}^{\perp }\right. \left\vert f_{u}^{\perp }\right\rangle =1$%
. By using Eq.~(\ref{mixedTerms}), we can always decompose $\left\vert
f_{u\oplus 1}\right\rangle =\cos x\left\vert f_{u}\right\rangle +e^{i\varphi
}\sin x\left\vert f_{u}^{\perp }\right\rangle $ with $\varphi $ arbitrary
phase, so that%
\begin{equation}
\rho _{F}=\left(
\begin{array}{cc}
\left\vert f_{u}\right\rangle & \left\vert f_{u}^{\perp }\right\rangle%
\end{array}%
\right) \left(
\begin{array}{cc}
\frac{1+\cos ^{2}x}{2} & \frac{e^{-i\varphi }\sin 2x}{^{{}}4_{{}}} \\
\frac{e^{i\varphi }\sin 2x}{^{{}}4_{{}}} & \frac{\sin ^{2}x}{2}%
\end{array}%
\right) \left(
\begin{array}{c}
\left\langle f_{u}\right\vert \\
\left\langle f_{u}^{\perp }\right\vert%
\end{array}%
\right) ~.
\end{equation}%
By means of a suitable unitary we then get%
\begin{equation}
\hat{U}\rho _{F}\hat{U}^{\dagger }=\lambda \left\vert \Phi _{-}\right\rangle
\left\langle \Phi _{-}\right\vert +(1-\lambda )\left\vert \Phi
_{+}\right\rangle \left\langle \Phi _{+}\right\vert ~,
\end{equation}%
where%
\begin{equation}
\lambda (x)=\frac{1-\left\vert \cos x\right\vert }{2}~,~\left\vert \Phi
_{\pm }\right\rangle =\frac{e^{-i\varphi }\left( 1+\cos 2x\pm 2\left\vert
\cos x\right\vert \right) \left( \csc 2x\right) \left\vert
f_{u}\right\rangle +\left\vert f_{u}^{\perp }\right\rangle }{N_{\pm }}~,
\label{Diagonal}
\end{equation}%
and $N_{\pm }^{2}=1+\left( 1+\cos 2x\pm 2\left\vert \cos x\right\vert
\right) ^{2}\left( \csc 2x\right) ^{2}$. Since $\left\langle \Phi
_{+}\right. \left\vert \Phi _{-}\right\rangle =0$, we simply achieve $S(\rho
_{F})=S(\hat{U}\rho _{F}\hat{U}^{\dagger })=H[\lambda (x)]$. In order to
compute the other entropy $S(\rho _{D})$, we just introduce an analogous
orthonormal set $\{\left\vert d_{u}\right\rangle ,\left\vert d_{u}^{\perp
}\right\rangle \}$ which leads to the corresponding result $S(\rho
_{D})=H[\lambda (y)]$.

Now, by setting $x=y$, we clearly have $S(\rho _{F})=S(\rho _{D})=H[\lambda
(x)]$. Then, we also have $\lambda (x)=D(x,x)$ for $-\pi /2\leq x\leq \pi /2$
and $\lambda (x)=1-D(x,x)$ for $\pi /2\leq x\leq 3\pi /2$, so that we can
always write $S(\rho _{F})=S(\rho _{D})=H(D)$. By replacing the latter
result in Eq.~(\ref{Extract}) we finally get $S(\rho _{E})=2H(D)$ which
leads to the rate of Eq.~(\ref{DWrateBB84}).$~\blacksquare $

\section{The six-state protocol and its symmetric eavesdropping}

In the BB84 protocol the signal states represent the four equidistant poles
lying on the equator of the Bloch sphere. In order to enhance the security,
one can then think to \emph{saturate} the sphere by including the
exploitation of the remaining two poles. This is done in the six-state
protocol \cite{sixstate} where also the basis $\{\left\vert R\right\rangle
,\left\vert L\right\rangle \}:=2^{-1/2}\{\left\vert 0\right\rangle
+i\left\vert 1\right\rangle ,\left\vert 0\right\rangle -i\left\vert
1\right\rangle \}$ of the third Pauli operator $Y=iXZ$ is exploited in both
Alice's random encoding and Bob's random decoding. The six-state protocol is
then formulated like the BB84 protocol except that now we have three bases $%
\{Z,X,Y\}$ and, therefore, six possible signal states $\{\left\vert
u\right\rangle ;~u=0,1,+,-,R,L\}$ encoding a logical qubit according to the
mapping $\bar{0}=\left\vert 0\right\rangle \vee \left\vert +\right\rangle
\vee \left\vert R\right\rangle $ and $\bar{1}=\left\vert 1\right\rangle \vee
\left\vert -\right\rangle \vee \left\vert L\right\rangle $.

Since the six-state protocol is a symmetric extension of the BB84 to the
third Pauli operator, we consider the same extension for the symmetric
attacks. This means that an arbitrary symmetric attack against the six-state
protocol is defined by Eqs.~(\ref{U_gen}) and~(\ref{symmetric}) where now $%
u=\{0,1,+,-,R,L\}$. The corresponding channel is again described by Eq.~(\ref%
{Symmetric_Channel}) which now corresponds to a uniform contraction by $F-D$
of all the Bloch sphere. It is trivial to check that a symmetric collective
attack against the six-state protocol is characterized by the same
Devetak-Winter rate of Eq.~(\ref{DWrateSimple}), exactly as before \cite%
{Rates}.

Let us construct a simple example for the explicit computation of this rate.
We can simplify the structure of the attack by imposing the conditions of
Eq.~(\ref{mixedTerms}) for all the bases, i.e., for $u=\{0,1,+,-,R,L\}$. All
these conditions imply now Eq.~(\ref{Fid}) together with $1+F\cos x-D\cos
y=2F$, which are simultaneously satisfied if and only if $y=\pi /2$. As a
consequence, we have%
\begin{equation}
D=\frac{1-\cos x}{2-\cos x}:=D(x)~,  \label{QBER_6state}
\end{equation}%
and the unitary interaction $\hat{U}$ is completely determined by a single
angle $x$. In particular, we can realize all the previous conditions by
choosing the ancillas of Eq.~(\ref{choice}) with $y=\pi /2$, i.e.,
\begin{equation}
\left\vert F_{0}\right\rangle =\left(
\begin{array}{c}
\sqrt{F} \\
0 \\
0 \\
0%
\end{array}%
\right) ~,~\left\vert D_{0}\right\rangle =\left(
\begin{array}{c}
0 \\
\sqrt{D} \\
0 \\
0%
\end{array}%
\right) ~,~\left\vert F_{1}\right\rangle =\left(
\begin{array}{c}
\sqrt{F}\cos x \\
0 \\
0 \\
\sqrt{F}\sin x%
\end{array}%
\right) ~,~\left\vert D_{1}\right\rangle =\left(
\begin{array}{c}
0 \\
0 \\
\sqrt{D} \\
0%
\end{array}%
\right) ~,  \label{choice2}
\end{equation}%
where also $\left\vert D_{0}\right\rangle $ and $\left\vert
D_{1}\right\rangle $ are orthogonal. Let us denote by $\mathcal{\tilde{S}}(x)
$ the symmetric collective attack specified by the interaction of Eq.~(\ref%
{choice2}). Then, it is easy to prove that $\mathcal{\tilde{S}}(x)$ has a
Devetak-Winter rate equal to%
\begin{equation}
R_{DW}=1+\frac{3D}{2}\log _{2}\frac{D}{2}+\left( 1-\frac{3D}{2}\right) \log
_{2}\left( 1-\frac{3D}{2}\right) ~,  \label{DWrateSixstates}
\end{equation}%
which corresponds exactly to the unconditionally-secure key-rate
of the
six-state protocol \cite{Lo2001} (with unconditional security threshold $%
D\simeq 12.6\%$).

\bigskip

\textbf{Proof of Eq.~(\ref{DWrateSixstates}).~}In order to get the result we
have to compute $S(\rho _{E})$ for the simple attack $\mathcal{\tilde{S}}(x)$%
. Eve's output state $\rho _{E}(u)$ has the same form of Eq.~(\ref{outputEVE}%
). Thus, the average state $\rho _{E}$ can be again recasted in terms of the
states $\rho _{D}$ and $\rho _{F}$ of Eq.~(\ref{rho_di}), in such a way to
represent the same quantum ensemble $\mathcal{\tilde{Q}}$ of Eq.~(\ref%
{Q_tilde}). As a consequence, the entropy $S(\rho _{E})$ can be again
extracted from of Eq.~(\ref{Extract}), where the computation of $S(\rho _{F})
$ and $S(\rho _{D})$ is now different since we have $y=\pi /2$ and not $x=y$
as before. The computation of $S(\rho _{D})$ is very easy thanks to the
orthogonality which now exists between the $D$'s states. Since $\left\langle
d_{u}\right. \left\vert d_{u\oplus 1}\right\rangle =\left\langle
D_{u}\right. \left\vert D_{u\oplus 1}\right\rangle =0$, we have in fact $%
S(\rho _{D})=H(1/2)=1$. The computation of $S(\rho _{F})$ is the same as
before except that now the eigenvalue $\lambda (x)$ of Eq.~(\ref{Diagonal})
is differently connected to the QBER $D(x)$ of Eq.~(\ref{QBER_6state}). It
is easy to check that $\lambda (x)=\left[ 1-D(x)\right] ^{-1}D(x)/2$ for $%
-\pi /2\leq x\leq \pi /2$ and $\lambda (x)=1-\{\left[ 1-D(x)\right]
^{-1}D(x)/2\}$ for $\pi /2\leq x\leq 3\pi /2$, so that we can always write $%
S(\rho _{F})=H(\lambda )=H\left[ (1-D)^{-1}D/2\right] $. By inserting the
latter result and $S(\rho _{D})=1$ into Eq.~(\ref{Extract}), one gets $%
S(\rho _{E})=D+H(D)+(1-D)H\left[ (1-D)^{-1}D/2\right] $ and, therefore, the
Devetak-Winter rate
\begin{equation}
R_{DW}=(1-D)\left\{ 1-H\left[ \frac{D}{2(1-D)}\right] \right\} -H(D)~,
\end{equation}%
which is equivalent to the result of Eq.~(\ref{DWrateSixstates}).~$%
\blacksquare $

\section{Conclusion}

In conclusion, we have considered very simple collective attacks
against the BB84 and six-state protocols, which are constructed by
trivially extending the individual symmetric attacks of Gisin
\textit{et al.} \cite{Symm}. Such symmetric collective attacks
have been proven to be sufficiently strong in order to minimize
the Devetak-Winter rates of these protocols. In fact, it has been
shown how to construct simple examples able to reach their
unconditionally-secure key-rates. Our results can be useful in the
cryptoanalysis of other QKD\ protocols which are based on the
symmetric exploitation of the vertices of regular polygons or
polyhedrons embedded in the Bloch sphere.

\section{Acknowledgements}

The research of the author was supported by a Marie Curie
International Fellowship within the 6th European Community
Framework Programme. The author thanks M. Lucamarini for
interesting discussions.


\begin{thebibliography}{99}
\bibitem{deFinetti} In particular, the quantum formulation of the classical
de Finetti theorem which is relevant here is the one concerning finite
exchangeability and finite-dimensional Hilbert spaces.
\bibitem{NoteDWrate} In general, the Devetak-Winter rate is defined for a $%
\{c\rightarrow qq\}$ channel where $R_{DW}:=\chi _{AB}-\chi _{AE}$. Here, we
consider its specialization to a $\{c\rightarrow cq\}$ channel, where Bob
performs incoherent decoding measurements.
\end{thebibliography}
\end{document}